\documentclass[a4paper]{article}
\usepackage[german]{babel}
\usepackage{amsmath,amssymb,amsthm}

\date{19.1.2009}

\newtheorem{folgerung}{Folgerung}
\newtheorem{lemma}{Lemma} 
\newtheorem{theorem}{Satz}

\def\cz{\mathbb{C}} 
\def\rz{\mathbb{R}} 

\def\const{\mathrm{C}} 

\newcommand{\cE}{\mathcal{E}}
\def\cI{\mathcal{I}}

\def\gp{\mathfrak{p}}
\def\gq{\mathfrak{q}}
\def\gx{\mathfrak{x}}
\def\gy{\mathfrak{y}}
\def\gS{\mathfrak{S}}

\def\rd{\mathrm{d}}
\def\ri{\mathrm{i}}
\def\sp{\mathop{\mathrm{Sp}}\nolimits} 

\def\ehf{E_\mathrm{HF}(Z)} 
\def\emu{E_\mathrm{M}(Z)} 
\def\erhf{E_\mathrm{rHF}(Z)} 
\def\es{E_\mathrm{S}(Z)} 
\def\htf{H_\mathrm{TF}} 

\begin{document}
\title{Das asymptotische Verhalten der
  Grundzustandsenergie des M\"ullerfunktionals f"ur schwere Atome\\
  English title: The asymptotic behaviour of the ground state energy
  of the M\"uller functional for heavy atoms}

\author{Heinz Siedentop\thanks{Mein Dank gilt dem Institute for
    Mathematics and its Applications an der University of Minnesota
    f\"ur seine gro\ss z\"ugige Unterst\"utzung meiner Forschung sowie
    der Deutschen Forschungsgemeinschaft, die diese Arbeit teilweise
    im Rahmen des SFB TR 12 gef"ordert hat. Mein besonderer Dank gilt
    Rupert Frank f"ur seine kritischen Bemerkungen zum Manuskript.}\\
  Mathematisches Institut\\
  Ludwig-Maximilians-Universit\"at M\"unchen\\
  Theresienstra\ss e 39\\ DE-80333 M\"unchen\thanks{Derzeitige
    Anschrift: Institute for Mathematics and its Applications,
    University of Minnesota, 114 Lind Hall, 207 Church Street S.E.,
    Minneapolis, MN 55455-0134,
    USA}\\
  Email: {\tt h.s@lmu.de} }

\maketitle

\begin{abstract}
  Wir zeigen, da"s die Grundzustandsenergie $E_\mathrm{M}(Z)$ des
  M"ullerfunktionals von (neutralen) Atomen der Ordnungszahl $Z$ mit
  der quantenmechanischen Grundzustandsenergie $\es$ bis zur
  Ordnung $o(Z^{5/3})$ "ubereinstimmt, d.h.
  $$
  \emu = \es + o(Z^{\tfrac53}).$$
  \end{abstract}


\section{Einleitung\label{sec:einleitung}}

\subsection{Quantenmechanik\label{sec:quant}}

Der Zustandsraum eines $N$-Elektronensystems sind die Einheitsstrahlen
im $N$-Teilchen-Hilbertraum
$$\mathfrak{H}_N:= \bigwedge_{n=1}^N L^2(\rz^3)\otimes \cz^2.$$
Der Schr"odingeroperator eines $N$-Elektronenatoms
\begin{equation} 
  \label{ham} 
  S_{N,Z}=\sum_{n=1}^N \left( -\Delta_n
    -\frac Z{|\gx_n|}\right)
  + \sum_{1\leq m<n\leq N}\frac1{|\gx_m-\gx_n|}
\end{equation}
ist in diesem Raum kanonisch als St"orung des Laplaceoperators
selbstadjungiert realisiert. Dessen Grunzustandsenergie ist
$E_\mathrm{S}(N,Z):=\inf(\sigma(S_{N,Z}))$. Wir vermerken f"ur
sp"ater, da"s wir im neutralen Fall, also $N=Z$, das erste Argument
aller auftauchenden Energiefunktionale einfach unterdr"ucken, d.h. wir
werden z.B. kurz $\es$ f"ur $E_\mathrm{S}(Z,Z)$ schreiben.

\subsection{Funktional der Einteilchendichtematrix\label{sec:funkt-der-eint}}

Wir nennen $\gamma$ eine (fermionische) Einteilchendichtematrix, wenn
$\gamma$ ein selbstadjungierter Operator der Spurklasse
$\gS^1(L^2(\rz^3)\otimes\cz^2)$ auf $L^2(\rz^3)\otimes\cz^2$ ist und
$0\leq\gamma\leq1$ gilt. Die Spur von $\gamma$ hei"st Teilchenzahl.
Wir bezeichnen die Menge der Einteilchendichtematrizes mit
\begin{equation}
  \label{eq:2}
  \cI := \{\gamma\in \gS^1(L^2(\rz^3)\otimes\cz^2)\ 
  |\ 0\leq \gamma \leq 1\}
\end{equation}
und setzen
\begin{equation}
\label{eq:3}
  \cI_N:=\{\gamma\in\cI\ |\ \sp\gamma\leq N\}.
\end{equation}

Nach Gilbert \cite{Gilbert1975} kann die Grundzustandsenergie eines
jeden quantenmechanischen Coulombsystems -- analog zum
Hohenberg-Kohn-Theorem -- als Minimum eines universellen Funktionals
der Einteilchendichtematrix $\gamma$ geschrieben werden.

Obwohl dieses Funktional unbekannt ist, gibt es mehrere Funktionale,
die eine gute Approximation f"ur $E_S(Z)$ liefern. Das bekannteste ist
wohl das Hartree-Fock-Funktional in der von Lieb \cite{Lieb1981V}
angegebenen Form. Um das Funktional einzuf"uhren, verwenden wir
folgende Notation: Zu gegebenem $\gamma\in\cI$ sei
$\varphi_1,\varphi_2,\dots$ ein vollst"andiges System orthonormierter
Eigenvektoren mit zugeh"origen Eigenwerten $\lambda_1,
\lambda_2,\dots$. Dann ist $\gamma(x,y):= \sum_{n=0}^\infty \lambda_n
\varphi_n(x)\overline{\varphi_n(y)}$ der Integralkern von $\gamma$ und
$\rho_\gamma(\gx):=\sum_{\sigma=1}^2 \gamma(x,x)$, die spinsummierte
Diagonale, die Teilchendichte von $\gamma$. (Wie nicht un"ublich,
bezeichnen wir mit $x$ und $y$ Raum-Spin-Variablen, also
$x,y\in\Gamma:=\rz^3\times\{1,2\}$ und sp"ater mit $\int_\Gamma\rd x$
die Integration bez"uglich $\gx$ und Summation "uber $\sigma$.)
Ferner sei
$$ D(f,g):= \frac12\int_{\rz^3}\rd\gx \int_{\rz^3}\rd\gy 
{\overline{f(\gx)}g(\gy) \over |\gx-\gy|}$$
das Coulombskalarprodukt und mit $X$ das Austausch-Skalarprodukt
$$X(\gamma,\delta):=\frac12 \int_\Gamma\rd x \int_\Gamma\rd y 
{\overline{\gamma(x,y)}\delta(x,y)\over |\gx-\gy|}. 
$$
(Die zugeh"origen quadratischen Formen bezeichnen wir mit denselben
Buchstaben, setzen jedoch das Argument in eckige Klammern.)

Damit k"onnen wir das Hartree-Fock-Funktional
\begin{eqnarray}
  \label{eq:1}
  \cE_\mathrm{HF}: \cI &\rightarrow & \rz \\
  \label{eq:4}
  \gamma&\mapsto& \sp[(-\Delta-Z/|\cdot|)\gamma] + D[\rho_\gamma]- X[\gamma]
\end{eqnarray}
definieren. Die Grundzustandsenergie des Hartree-Fock-Funktionals,
kurz die Har\-tree-Fock-Energie, ist
\begin{equation}
  \label{eq:5}
  E_\mathrm{HF}(N,Z) = \inf \{\cE_\mathrm{HF}(\gamma)\ |\ \gamma\in\cI_N\}.
\end{equation}
Es ist bekannt, da"s das Infimum in \eqref{eq:5} f"ur ganzzahliges $N$
unver"andert bleibt, wenn man die Minimierung auf idempotente
Dichtematrizes, also solche, die von Slaterdeterminanten herr"uhren,
beschr"ankt (Lieb \cite{Lieb1981V}).  Ferner ist bekannt, da"s das
Infimum f"ur $N<Z+1$ stets f"ur eine Dichtematrix mit Teilchenzahl $N$
angenommen wird, d.h. ein Grundzustand mit Teilchenzahl $N$ existiert
(Lieb und Simon \cite{LiebSimon1974,LiebSimon1977T}).

Da f"ur idempotentes $\gamma=|\varphi_1\rangle\langle\varphi_1|
+\cdots+|\varphi_N\rangle\langle\varphi_N|$ mit orthonormalen Spinoren
$\varphi_1,\cdots,\varphi_N$ die Identit"at
$$\cE_\mathrm{HF}(\gamma) = N!^{-1}(\varphi_1\wedge\cdots\wedge\varphi_N, 
S_{N,Z}\varphi_1\wedge\cdots\wedge\varphi_N)$$ gilt, ist die
Hartree-Fock-Energie $E_S(N,Z)$ stets eine obere Schranke an die
quantenmechanische Grundzustandsenergie.

Ein weiteres Funktional der Einteilchendichtematrix wurde von M"uller
\cite{Muller1984} angegeben
\begin{eqnarray}
  \label{eq:6}
  \cE_\mathrm{M}: \cI &\rightarrow & \rz \\
  \label{eq:7}
  \gamma&\mapsto& \sp[(-\Delta-Z/|\cdot|)\gamma] + D[\rho_\gamma]
- X[\gamma^{\tfrac12}]
\end{eqnarray}
Die M"ullerenergie ist
\begin{equation}
  \label{eq:8}
  E_\mathrm{M}(N,Z) = \inf \{\cE_\mathrm{M}(\gamma)\ |\ \gamma\in\cI_N\}.
\end{equation}
Frank u.a. \cite{Franketal2007} zeigten, da"s f"ur $N\leq Z$ der --
auch ohne diese Einschr"ankung existierende Minimierer -- stets die
Teilchenzahl $N$ besitzt, d.h. es exisitiert ein Grundzustand mit
Teilchenzahl $N$. Im Gegensatz zu einem Hartree-Fock-Minimierer
$\gamma_\mathrm{HF}$, dessen Bildraum stets $N$ dimensional ist, ist
der Bildraum eines M"ullerminimierers jedoch immer
unendlich-dimensional.  Offensichtlich ist, wiederum per
constructionem, da"s die M"ullerenergie kleiner als die
Hartree-Fock-Energie ist. Weiter gilt, da"s im Zweielektronenfall
(auch mehrere Zentren sind erlaubt), die M"ullerenergie sogar die
quantenmechanische Energie nach unten beschr"ankt
\cite{Franketal2007}, ein Faktum, da"s durch numerische Rechnungen
f"ur alle Elektronenzahlen $N$ nahe gelegt wird; mehr noch, diese
numerischen Resultate deuten darauf hin, da"s die M"ullerenergien
nicht mehr von den quantenmechanischen Energien abweichen als die
Hartree-Fock-Energien. W"ahrend jedoch auf der analytischen Seite f"ur
die Hartree-Fock-Energie
\begin{equation}
  \label{eq:9}
  \ehf = \es + o(Z^{\tfrac53})
\end{equation}
bekannt ist (Bach \cite{Bach1993}) -- dieses Resultat gilt auch f"ur
Molek"ule und Ionen, wenn die Neutralit"at des Systems nicht zu stark
verletzt ist --, fehlt ein entsprechendes Resultat f"ur die
M"ullerenergie. Es k"onnte also sein, da"s die M"ullerenergie
unrealistisch klein ist. Da"s dieses nicht zutrifft, ist der Inhalt
der vorliegenden Arbeit. Genauer gesagt ist unser Resultat
\begin{theorem}
  \label{t1}
  Es gibt eine positive Konstante $c$ so, da"s f"ur alle $Z\geq1$
  $$\emu \leq \ehf \leq \emu + c Z^{\tfrac53-\tfrac1{140}}$$
gilt.
\end{theorem}
Wir bemerken, da"s dieses Resultat auf Molek"ule mit nicht zu
kleinen Kernabst"anden -- einschlie"slich der physikalischen Abst"ande
-- "ubertragbar ist. Ferner ist die strikte Neutralit"at nicht
wesentlich. Das System mu"s nur im wesentlichen neutral sein. Wir
pr"asentieren jedoch der "Ubersichtlichkeit halber hier nur den
atomaren Fall.

Weiterhin sei bemerkt, da"s eine durch numerische Resultate nahe
gelegte Ungleichung $\es\geq \emu$ auch sofort das Bachsche Resultat
\eqref{eq:9} implizieren w"urde.

Desweiteren m"ochten wir darauf verweisen, da"s erst j"ungst
Dichtefunktionale hergeleitet und untersucht wurden, die die gleiche
asymptotische Genauigkeit (bis zur Ordnung $o(Z^{5/3})$) besitzen, wie
das M"ullerfunktional (Lee u. a. \cite{Leeetal2008}). W"ahrend die
Konstruktion solcher Funktionale erheblichen Aufwand erfordert,
ist dieses im M"ullerfunktional ohne weiteres eingebaut.

Weiter bemerken wir, da"s Satz \ref{t1} g"ultig bleibt, wenn man im
M"ullerfunktional $X[\gamma^{1/2}]$ durch $X[\gamma^\alpha]$ mit
$1/2\leq\alpha\leq1$ ersetzt. Solche Funktionale wurden von
Sharma u.a. \cite{Sharmaetal2008} eingef"uhrt.

Schlie"slich impliziert der Satz \ref{t1} mittels des obigen Bachschen
Resultats und der asymptotischen Entwicklung der quantenmechanischen
Grundszustandsenergie (Fefferman und Seco
\cite{FeffermanSeco1989,FeffermanSeco1990O,FeffermanSeco1992,FeffermanSeco1993,FeffermanSeco1994,FeffermanSeco1994T,FeffermanSeco1994Th,FeffermanSeco1995})
sofort, da"s
\begin{equation}
  \label{eq:10}
  \emu = \es + o(Z^{\tfrac53})=E_\mathrm{TF}(Z)+ \frac14Z^2 
- c_\mathrm{S}Z^{\tfrac53} +o(Z^{\tfrac53})
\end{equation}
gilt. Hier ist $E_\mathrm{TF}(Z)$ die Thomas-Fermi-Energie des
neutralen Atoms der Ordnungszahl $Z$ und $c_\mathrm{S}$ ist die
Schwingerkonstante (Schwinger \cite{Schwinger1981}). 

Der Beweis des Satzes \ref{t1} ist von der Bachschen Idee
\cite{Bach1992} gepr"agt, die Energiedifferenz zwischen zwei
Funktionalen durch eine Norm der abgeschnittenen
Einteilchendichtematrix $\gamma(1-\gamma)$ abzusch"atzen und diese
Norm dann als klein zu erkennen. Dieses wiederum wird durch Vergleich
mit einem geeigneten approximativen Zustand, hier dem halbklassichen
Grundzustand des Schr"odingeroperators mit Thomas-Fermi-Potential,
gewonnen, wobei unser Vorgehen von Graf und Solovej
\cite{GrafSolovej1994} inspiriert ist.

\section{Einfache Schranke an die M\"ullerenergie}
\label{sec:einfach}

Wir ben"otigen eine einfache Schranke der Gr"o"senordnung $Z^{\tfrac53}$ an
die M"ullerenergie, um die Asymptotik in dieser Gr"o"senordnung zu
finden. Eine solche Schranke wird in diesem Abschnitt entwickelt.

Cioslowski und Pernal \cite{CioslowskiPernal1999} (siehe auch
\cite{Franketal2007}) konnten zeigen, da"s der M"ullersche
Austausch-Korrelations-Term keine einfache Ungleichung vom
Lieb-Oxford-Typ (Lieb \cite{Lieb1979}, Lieb und Oxford
\cite{LiebOxford1981}) der Form $X[\gamma^{\tfrac12}]\leq \const
\int\rho_\gamma^{\tfrac43}$ zul"a"st.\footnote{Hier und im
  folgenden bezeichne $\const$ eine generische Konstante.} Dennoch
bewirkt dieser Term -- wie der Austauschterm des
Hartree-Fock-Funk\-tio\-nals -- eine Energieerniedrigung, die
h"ochstens der Gr"o"senordnung $Z^{\tfrac53}$ ist, was aus folgendem
Lemma folgt; denn das reduzierte Hartree-Fock-Funktional
\begin{equation}
    \label{eq:12}
    \cE_\mathrm{rHF}(\gamma) :=  \sp[(-\Delta-Z/|\cdot|)\gamma] + D[\rho_\gamma]
  \end{equation}
  ist offensichtlich gr"o"ser als das Hartree-Fock-Funktional, da ein
  negativer Term entfallen ist. Ferner sei $E_\mathrm{rHF}(N,Z)$ 
  das Infimum von $\cE_\mathrm{rHF}$ auf $\cI_N$.
\begin{lemma}
  \label{l1}
  Sei $\gamma$ ein Minimierer des M"ullerfunktionals auf $\cI_N$. Dann 
  gilt 
  \begin{equation}
    \label{eq:12aa}
    \emu\leq \ehf \leq \erhf\leq \cE_\mathrm{rHF}(\gamma) 
  \leq \emu + \const Z^{\tfrac53}.
  \end{equation}
\end{lemma}
\begin{proof}
  Die ersten drei Ungleichungen sind per constructionem und
  Variationsprinzip offensichtlich richtig, soda"s wir lediglich die
  letzte Schranke zeigen. Dazu w"ahlen wir ein sp"ater n"aher zu
  bestimmendes $\epsilon\in(0,1)$ und bezeichnen mit $\gamma$ einen
  Minimierer des M"ullerfunktionals. Dann gilt
  \begin{multline}
    \label{eq:11}
    E_\mathrm{rHF}(Z,Z)\geq E_\mathrm{HF}(Z,Z)\geq E_\mathrm{M}(Z,Z) =
    \sp[(-\Delta-Z/|\cdot|)\gamma] + D[\rho_\gamma]
    - X[\gamma^{\tfrac12}] \\
    \geq \sp[(-(1-\epsilon)\Delta-Z/|\cdot|)\gamma] + D[\rho_\gamma] +
    \epsilon \sp(-\Delta\gamma)
    - X[\gamma^{\tfrac12}]\\
    \geq {1\over1-\epsilon}E_\mathrm{rHF}(Z) +
    \inf\Big\{\int_\Gamma\rd x
    \underbrace{\int_\Gamma\epsilon|\nabla_\gy\gamma^{\tfrac12}(x,y)|^2
      -\frac{|\gamma^{\tfrac12}(x,y)|^2}{2|\gx-\gy|}\rd y}_{\geq - \tfrac1{16\epsilon}\int_\Gamma |\gamma^{\tfrac12}(x,y)|^2\rd y}  \big| \gamma\in \cI_N\Big\}\\
    \geq E_\mathrm{rHF}(Z)-\epsilon\const Z^{\tfrac73} - \epsilon^{-1}
    N/16 = E_\mathrm{rHF}(Z) - \const Z^{\tfrac53}
  \end{multline}
  Hier haben wir neben der Skalierung des reduzierten
  Hartree-Fock-Funktionals benutzt, da"s $E_\mathrm{rHF}(Z) =
  E_\mathrm{TF}(1)Z^{\tfrac73} + o(Z^{\tfrac73})$ gilt, was
  aus dem Beweis von \cite[Satz 5.1]{Lieb1981} folgt. Sodann beachten
  wir, da"s $N=Z$ gilt und wir deshalb $\epsilon= Z^{-\tfrac23}$ w"ahlen.
\end{proof}
\begin{lemma}[Virialsatz (Frank u.a \cite{Franketal2007})]
  \label{l2}
  Sei $\gamma$ ein Minimierer des M"ullerfunktionals $\cE_M$ auf
  $\cI_N$. Dann gilt $\sp(-\Delta\gamma)=-\cE_M(\gamma)$.
\end{lemma}
\begin{proof}
  Wir strecken den Grundzustand $\gamma$ und definieren f"ur positives
  $\lambda$ die gestreckte Dichtematrix $\gamma_\lambda$ durch
  $\gamma_\lambda(x,y):=\lambda^3\gamma(\lambda\gx,\sigma;\lambda\gy,\tau)$.
  Die gestreckte Dichtematrix geh"ort offensichtlich auch zu $\cI_N$.
  Da $f(\lambda):=\cE_M(\gamma_\lambda)$ f"ur positives Argument
  differenzierbar ist und f"ur $\lambda=1$ ein Minimum hat, folgt die
  Aussage aus $f'(1)=0$.
\end{proof}

Da $\ehf= E_\mathrm{TF}(1) Z^{\tfrac73} + o(Z^{\tfrac73})$ (Lieb
und Simon \cite{LiebSimon1977}) gilt, folgt aus Lemmata \ref{l1} und
\ref{l2}, da"s
 \begin{equation}
   \label{eq:12a}
   \sp(-\Delta\gamma) = O(Z^{\tfrac73})
 \end{equation}
gilt.

\section{Verdampfungsgrad des M\"ullergrundzustandes\label{sec:2a}}
Der Bildraum einer das M"ullerfunktional minimierenden
Einteilchendichtematrix $\gamma$ ist unendlichdimensional (Frank u.a.
\cite{Franketal2007}), eine Eigenschaft, die das M"ullerfunktional mit
der Schr"odingertheorie teil (Friesecke \cite{Friesecke2003}, Lewin
\cite{Lewin2004}). Frank u.a. vermuten sogar, da"s ein solches
$\gamma$ trivialen Kern hat. Von der Einteilchendichtematrix eines
quantenmechanischen Grundzustandes ist seit Bach \cite{Bach1992}
bekannt, da"s sie nicht zu stark von einer Projektion
abweichen kann. Ein Ma"s daf"ur ist die Spur der Dichtematrix
$\gamma(1-\gamma)$, die verschwindet, wenn $\gamma$ eine Projektion ist. Eine st"arkere
Eigenschaft ist die fast vollst"andige Kondensation der Dichtematrix
in einen vorgegebenen Zustand, eine Eigenschaft, die im
quantenmechanischen Fall von Graf und Solovej \cite{GrafSolovej1994}
untersucht wurde. Praktikabilit"ats halber w"ahlen wir die
halbklassiche Projektion $P$ auf die gebundenen Zust"ande im
Thomas-Fermi-Potential: Sei $\phi:= Z/|\cdot| - \rho*|\cdot|^{-1}$ das
Thomas-Fermi-Potential, das durch den Minimierer $\rho$ des
Thomas-Fermi-Funktionals $\cE_\mathrm{TF}$ eines Atoms mit Ladungszahl
$Z$ gegeben ist. Ferner sei
$$g(\gx):=
\begin{cases}
  (2\pi R)^{-\tfrac12}|\gx|^{-1}\sin(\pi|\gx|/R)& |\gx|\leq R\\
  0& |\gx|>R
\end{cases}
$$ 
mit $R=Z^{-3/5}$.  Weiter f"uhren wir den koh"arenten Zustand mit
Implus $\gp$, Ort $\gq$ und Spin $\sigma$ ein, d.h.
\begin{equation}
  \label{eq:12b}
  |\alpha\rangle=|\gp,\gq,\tau\rangle := e^{\ri\gp\cdot\cdot}g(\cdot-\gq) \delta_{\cdot,\sigma}
\end{equation}
mit $\alpha:=(\gp,\gq,\sigma)\in \Gamma:=
\rz^3\times\rz^3\times\{1,2\}$, und
$$
\int_\Gamma \rm d \Omega(\alpha) := (2\pi)^{-3} \int_{\rz^3}\rd \gp
\int_{\rz^3}\rd \gq \sum_{\tau=1,2}
$$ 
sei das nat"urliche Ma"s auf $\Gamma$ im Sinne von Planck, das die
Elektronenzust"ande im Phasenraum z"ahlt. Ferner sei $M$ der klassisch
erlaubte Bereich des Phasenraumes, d.h.
$$ M := \{(\gp,\gq,\sigma)\in \Gamma\ |\ \gp^2 - \phi(\gq)<0\}.$$
Der "`halbklassische Projektor"'\footnote{Es sei vermerkt, da"s der
  so eingef"uhrte Operator $P$ kein Projektor ist.} des
Schr"odingeroperators mit Thomas-Fermi-Po\-ten\-tial ist
\begin{equation}
  \label{eq:12c}
  P=  \int_M \rd \Omega(\alpha)|\alpha\rangle\langle\alpha|.
\end{equation}
Damit haben wir
\begin{lemma}
  Sei $\gamma$ ein Grundzustand des M"ullerfunktionals eines neutralen
  Atoms der Ordnungszahl $Z$. Dann gilt f"ur den Verdampfungsgrad
  $\delta(\gamma,P)$ des Zustands $\gamma$ aus dem fermionischen
  Zustand $P$
  \label{l3}
  \begin{equation}
    \label{eq:12ca}
    \delta(\gamma,P):=\sp((1-P)\gamma) = O(Z^{\tfrac{69}{70}}).
  \end{equation}
\end{lemma}
\begin{proof}
  Zun"achst erinnern wir daran, da"s f"ur alle $\gamma\in\cI$
  \begin{equation}
    \label{eq:12d}
    \sp(\gamma(\underbrace{-\Delta-\phi}_{=:\htf})) \geq \int_\Gamma\rd\Omega(\alpha) (\gp^2-\phi(\gx))_- - \const Z^{\tfrac73-\tfrac1{30}}
  \end{equation}
  gilt (Lieb [Abschnitt V.A.2]\cite{Lieb1981}, Thirring
  \cite{Thirring1981}). Damit haben wir
  \begin{equation}
    \label{eq:12e}
    \cE_\mathrm{rHF}(\gamma) \geq \sp(\gamma \htf) -D[\rho]\geq \int_\Gamma\rd
    \Omega(\alpha) (\gp^2-\phi(\gx))_- - D[\rho] - \const Z^{\tfrac73-\tfrac1{30}} ,
  \end{equation}
  da $D[\rho-\rho_{\gamma_\mathrm{rHF}}]\geq0$.  Desweiteren benutzen
  wir die Phasenraumabsch"atzung (Bach \cite[Lemma 12]{Bach1993}, Graf
  und Solovej \cite[Lemma 7 und den Beweis von
  (1.16)]{GrafSolovej1994})
  \begin{equation}
    \label{eq:12f}
    \int_{-a<\gp^2-\phi(\gx)<0}\rd\Omega(\alpha)(\gp^2-\phi(\gx)+a) 
    = O( a^{\tfrac74}).
  \end{equation}
  Der Vollst"andigkeit halber f"ugen wir den Beweis der Absch"atzung
  \eqref{eq:12f} von Graf und Solovej in leicht vereinfachter Form
  hier ein:\begin{multline}
    \label{eq:17}
    \int_{-a<\gp^2-\phi(\gx)<0}\rd\Omega(\alpha)(\gp^2-\phi(\gx)+a)
    = \int_0^a\rd b \int_{-b<\gp^2-\phi(\gx)<0}\rd\Omega(\alpha)\\
    = \frac{4\pi}3 \int_0^a\rd b
    \int_{\rz^3}\rd\gx(\phi(\gx)^{\tfrac32} -
    (\phi(\gx)-b)_+^{\tfrac32}) = 2\pi \int_0^a\rd b
    \int^b_0\rd c \int_{\rz^3}(\phi(\gx)-c)_+^{\tfrac12}\\
    \leq 2\pi\int_0^a\rd b\int^b_0\rd c
    \int_{\rz^3}\rd\gx(\const|\gx|^{-4}-c)_+^{\tfrac12} =\const
    \int_0^a\rd b\int^b_0\rd c c^{-\tfrac14}=\const\ c^{\tfrac74},
  \end{multline}
  wo wir das Potential $\phi$ durch die Sommerfeldl"osung $\const
  |\gx|^{-4}$ nach oben abgesch"atzt haben (Lieb \cite[S.
  607]{Lieb1981}).

  Weiter sei
  $$\cE_{\mathrm{M},a}(\gamma) := \cE_\mathrm{M}(\gamma)- a \sp(1-P)\gamma.$$
  Nun gilt
  \begin{multline}
    \label{eq:12g}
    a \sp (1-P)\gamma = \emu - \cE_{\mathrm{M},a}(\gamma)
    \underbrace{\leq}_{(13)} \erhf -  [\cE_\mathrm{rHF}(\gamma) - a \sp((1-P)\gamma)] +\const Z^{\tfrac53}\\
    \leq\erhf + aZ - \sp((\htf+aP)\gamma) +D[\rho] + \const Z^{\tfrac53}\\
    \leq \int_\Gamma\rd\Omega(\alpha) (\gp^2-\phi(\gx)+a)_- - D[\rho]
    - \int_\Gamma\rd\Omega(\alpha) (\gp^2-\phi(\gx) + a\chi_M(\gx))_-
    + D[\rho] + \const
    Z^{\tfrac{69}{30}}\\
    = \int_{-a<\gp^2-\phi(\gx)<0} \rd\Omega(\alpha)(\gp^2-\phi(\gx)+a)
    + O(Z^{\tfrac{69}{30}})\leq
    \const(a^{\tfrac74}+Z^{\tfrac{69}{30}}),
  \end{multline}
  d.h. f"ur $a= Z^{\tfrac{138}{105}}$ erhalten wir
  \begin{equation}
    \label{eq:12h}
    \delta(\gamma,P)= O(Z^{\tfrac{69}{70}}), 
  \end{equation} 
  also die behauptete Absch"atzung an den Verdampfungsgrad.
\end{proof}
Wir m"ochten an dieser Stelle keine Vermutung "uber den tats"achlichen
Verdampfungsgrad abgeben. Es ist aber klar, da"s die hergeleitete
Fehlerabsch"atzung nicht optimal ist. Benutzt man z.B. den
$Z$-Teilchen-Grundzustand von $\htf$ und die von Fefferman und Seco
\cite{FeffermanSeco1989,FeffermanSeco1990O,FeffermanSeco1992,FeffermanSeco1993,FeffermanSeco1994,FeffermanSeco1994T,FeffermanSeco1994Th,FeffermanSeco1995}
angegebene asymptotische Entwicklung, so ergibt sich ein Fehler der
maximalen Gr"o"se $O(Z^{\tfrac57})$. Wir halten also fest, da"s
\eqref{eq:12ca} zwar nicht optimal aber einfach und f"ur unsere Zwecke
hinreichend ist.
\begin{folgerung}
  \label{f1}
  Sei $\gamma$ ein Minimierer des M"ullerfunktionals auf $\cI_Z$. Dann
  ist
  \begin{equation}
    \label{eq:12i}
    \sp(\gamma(1-\gamma)) \leq \const Z^{\tfrac{69}{70}}.
  \end{equation}
\end{folgerung}
\begin{proof}
  Da $0\leq\gamma\leq1$ und $\sp\gamma=\sp P=Z$ ist, haben wir
  \begin{multline}
    \label{eq:12j}
    \sp(\gamma(1-\gamma)) = \sp(P\gamma(1-\gamma)) +
    \sp((1-P)\gamma(1-\gamma)) \leq \sp(P(1-\gamma)) + \sp((1-P)\gamma)\\
    = Z - \sp(P\gamma) + \sp((1-P)\gamma) = 2 \sp((1-P)\gamma) =
    2\delta(\gamma,P) = O(Z^{\tfrac{69}{70}}).
  \end{multline}
\end{proof}

\section{Schranke an die M\"ullerenergie: Abgeschnittene Einteilchendichtematrix}
\label{sec:3}
\begin{lemma}
  \label{l4}
  Sei $\gamma$ ein Minimierer des M"ullerfunktionals $\cE_\mathrm{M}$
  auf $\cI_Z$. Dann gilt 
  $$0\leq \ehf-\emu \leq \const Z^{\tfrac76}(\sp[\gamma(1-\gamma)])^{\tfrac12}.$$
\end{lemma}
\begin{proof}
    Gem"a"s Variationsprinzip haben wir
  \begin{align}
    \label{eq:13}
    0&\leq \ehf  -\emu \leq \frac12 \int_\Gamma\rd x
    \int_\Gamma\rd y
    {|\gamma^{\tfrac12}(x,y)|^2 - |\gamma(x,y)|^2 \over|\gx-\gy|}\\
    \label{eq:14}
    &= \frac12 \int_\Gamma\rd x \int_\Gamma\rd y
    {(\gamma^{\tfrac12}(x,y)+\gamma(x,y))\overline{(\gamma^{\tfrac12}(x,y) - \gamma(x,y))} \over|\gx-\gy|}\\
    \label{eq:15}
    &\leq \frac12 \sqrt{\int_\Gamma\rd x \int_\Gamma\rd y
      {|\gamma^{\tfrac12}(x,y) + \gamma(x,y)|^2\over|\gx-\gy|^2}} \sqrt{\sp|\gamma^{\tfrac12}(1-\gamma^{\tfrac12})|^2} \\
    \label{eq:16}
    &\leq 2 \sqrt{\sp(-\Delta\gamma)} \sqrt{\sp[\gamma(1-\gamma)]}\leq \const
    Z^{\tfrac76}\sqrt{\sp[\gamma(1-\gamma)]},
  \end{align}
  wo wir die Schwarzsche Ungleichung von \eqref{eq:14} nach
  \eqref{eq:15} und die Hardysche Ungleichung und
  $\gamma^{\tfrac12}\geq\gamma$ von \eqref{eq:15} nach \eqref{eq:16} sowie
  in der letzten Zeile $\sp(-\Delta\gamma)=O(Z^{\tfrac73})$ (Gleichung
  \eqref{eq:12a}) benutzt haben.
\end{proof}

\begin{proof}[Beweis des Satzes \ref{t1}]
  Die untere Schranke ist trivial; denn die Hartree-Fock-Energie ist
  stets gr"o"ser als die M"ullerenergie. Um die obere Schranke zu
  zeigen, m"ussen wir nun lediglich noch das Lemma \ref{l4} und die
  Folgerung \ref{f1} kombinieren und erhalten die behauptete
  Ungleichung des \ref{t1}. Satzes.
\end{proof}

\bibliography{coulomb}

\end{document}